\documentclass[12pt]{article}
\usepackage[english]{babel}
\usepackage[pdftex]{color}
\usepackage{amsmath}
\usepackage{graphicx}
\usepackage{hyperref}
\usepackage{amsmath}
\usepackage{amsfonts}
\usepackage{amssymb}

\begin{document}

\title{ \textbf{{} Comparative Analysis of Finite Field-dependent BRST
Transformations\thanks{%
Talk presented at SQS'15, 03 August --\ 08 August, 2015, at JINR, Dubna,
Russia}}}
\author{\textsc{Pavel Yu. Moshin and Alexander A. Reshetnyak\thanks{%
moshin@phys.tsu.ru \quad reshet@ispms.tsc.ru}} \\
%EndAName
Tomsk State University, Department of Physics, 634050, Tomsk, Russia,\\
Institute of Strength Physics and Materials Science, 634021, Tomsk, Russia}
\date{}
\maketitle

\begin{abstract}
We present a review of our recent study \cite{Reshetnyak,MR1,MR2,MR3,MR4,MR5}%
, %(A. Reshetnyak, IJMPA 29 (2014) 1450128; P. Moshin, A. Reshetnyak, Nucl.
%Phys. B 888 (2014) 92; Phys. Lett B 739 (2014) 110; IJMPA 29 (2014) 1450159;
%IJMPA 30 (2015) 1550021; [arxiv:1506.04660 [hep-th]),
in which the concept of finite field-dependent BRST and BRST-antiBRST
transformations for gauge theories was introduced, and their properties
investigated. An algorithm of exact calculation for the Jacobian of a
respective change of variables in the path integral is presented.
Applications to the Yang--Mills theory and Standard Model, in view of
infra-red (Gribov) peculiarities, are discussed.
\end{abstract}

\section{Introduction}

BRST transformations \cite{BRST1,BRST2} for gauge theories in Lagrangian
formalism were first examined in the capacity of \emph{field-dependent}
\emph{(FD)} BRST transformations within the field-antifield approach \cite%
{BV} in order to prove the independence from small gauge variations
(expressed through the gauge fermion $\psi $) of the path integral $Z_{\psi
} $: $Z_{\psi }=Z_{\psi +\delta \psi }$, with the choice $\mu =-\frac{\imath
}{\hbar }\delta \psi $ for the Grassmann-odd parameter of FD BRST
transformations. Originally introduced as the case of a special $N=1$\ SUSY
transformation, being a change of the field variables $\phi ^{A}$,
\begin{equation}
\phi ^{A}\rightarrow \phi ^{A\prime }=\phi ^{A}+\delta _{\mu }\phi ^{A}\ ,\
\ \mathcal{I}_{\phi }^{\psi }=d\phi \exp \Big\{\frac{\imath }{\hbar }S_{\psi
}(\phi )\Big\}\ ,\ \ Z_{\psi }=\int \mathcal{I}_{\phi }^{\psi },
\label{integrans}
\end{equation}%
in the\ integrand$\ \mathcal{I}_{\phi }^{\psi }$ with a quantum action $%
S_{\psi }(\phi )$, BRST transformations were extended, by means of antiBRST
transformations \cite{aBRST1,aBRST2} in Yang--Mills theories, to $N=2$
BRST-antiBRST transformations (in Yang--Mills \cite{aBRST4} and general
gauge theories \cite{BLT1}), which were associated with an $\mathrm{Sp}(2)$%
-doublet of Grassmann-odd parameters, $\mu _{a}$, $a=1,2$.

The concept of \emph{finite} FD BRST transformations was introduced by
Joglekar and Mandal \cite{JM} in Yang--Mills theories, as a sequence of
infinitesimal FD BRST transformations (with a numeric parameter $\kappa $),
in order to prove the gauge-independence of the path integral within the
family of $R_{\xi }$-gauges and their non-linear deformations in the field
variables.

The study \cite{Sorella} by a group of Brazilian researchers (see also the
references therein) suggested an analysis of so-called \emph{soft BRST
symmetry breaking} in Yang--Mills theories, with reference to the Gribov
problem \cite{Gribov} in the long-wave spectra of field configurations,
which also involves the Zwanziger proposal \cite{Zwanziger} for a horizon
functional joined additively to a BRST invariant quantum action. The study
of \cite{Sorella}, with the related scope of problems, caught the attention
of one the authors of the present article (A.A.R.), who subsequently (March,
2011) turned it to the attention of P.M.~Lavrov and then O.~Lechtenfeld. The
resulting study \cite{llr1} of this problem within the field-antifield
formalism suggested an equation for the BRST non-invariant addition $M(\phi
,\phi ^{\ast })$ to the quantum action $S_{\psi }(\phi ,\phi ^{\ast })$ of a
general gauge theory. The validity of this equation preserves the
gauge-independence of the vacuum functional $Z_{\psi ,M}(0)$ [see (\ref%
{genfun}) for a definition] and effective action, depending on external
antifields, $\Gamma _{M}=\Gamma _{M}(\phi ,\phi ^{\ast })$, and evaluated on
the extremals $\Gamma _{M}\frac{\overleftarrow{\delta }}{\delta \phi ^{A}}%
=\Gamma _{M}\overleftarrow{\partial }_{A}=\Gamma _{M,A}=0$,
\begin{eqnarray}
\hspace{-0.7em} &\hspace{-0.7em}&\hspace{-0.7em}\Big[M_{A}(\textstyle\frac{%
\hbar }{\imath }\overrightarrow{\partial }_{(J)},\phi ^{\ast })\Big(\hspace{%
-0.1em}\overrightarrow{\partial }{}^{\ast A}-\frac{\imath }{\hbar }M^{\ast
A}(\frac{\hbar }{\imath }\overrightarrow{\partial }_{J},\phi ^{\ast })%
\hspace{-0.1em}\Big)\delta \psi ({\textstyle\frac{\hbar }{\imath }}%
\overrightarrow{\partial }_{(J)})+\delta M(\frac{\hbar }{\imath }%
\overrightarrow{\partial }_{(J)},\phi ^{\ast })\hspace{-0.1em}\Big]Z_{\psi
,M}(J,\phi ^{\ast })\hspace{-0.1em}=\hspace{-0.1em}0  \label{softbreak} \\
\hspace{-0.7em} &\hspace{-0.7em}&\hspace{-0.7em}\Longrightarrow Z_{\psi
,M}(0)=Z_{\psi +\delta \psi ,M+\delta M}(0)\ \mathrm{and}\ \delta \Gamma _{M}%
\big|_{\Gamma _{M,A}=0}=0,\ \mathrm{where}\ \Gamma _{M}=\textstyle\frac{%
\hbar }{\imath }\ln Z-J_{A}\phi ^{A},  \label{softbreak1}
\end{eqnarray}%
where it is assumed that $[M_{A},M^{\ast A}]\equiv \lbrack M\overleftarrow{%
\partial }_{A},\overrightarrow{\partial }^{\ast A}M]$, and $\phi ^{A}=\frac{%
\hbar }{\imath }\overrightarrow{\partial }_{(J)}^{A}\ln Z$ are average
fields, having the same form when the horizon functional $H(A)$ for YM
fields $A^{\mu n}(x)$ is used as $M(\phi ,\phi ^{\ast })$. In terms of the
vacuum expectation value, in the presence of external sources $J_{A}$, and
with a given gauge $\psi $, relation (\ref{softbreak}) acquires the form,%
\footnote{%
In fact, the horizon functional in the family of $R_{\xi }$-gauges for small
$\xi $ was found explicitly in \cite{llr1}, see Eq.~(5.20) therein, by using
FD BRST transformations with a small odd-valued parameter.} $\frac{%
\overrightarrow{\delta }S_{\psi }}{\delta \phi _{A}^{\ast }}\equiv
\overrightarrow{\partial }{}^{\ast A}S_{\psi }$,%
\begin{equation}
\hspace{-0.25em}\hspace{-0.25em}\hspace{-0.25em}\left\langle \hspace{-0.1em}%
\delta M+M\overleftarrow{s}\textstyle\frac{\imath }{\hbar }\delta \psi (\phi
)\hspace{-0.1em}\right\rangle =\left\langle \hspace{-0.1em}\delta M-M%
\overleftarrow{s}\mu (\delta \psi )\hspace{-0.15em}\right\rangle \hspace{%
-0.1em}=\hspace{-0.1em}0,\ \mathrm{where}\,\overleftarrow{s}\hspace{-0.2em}=%
\hspace{-0.1em}\overleftarrow{\partial }_{A}\hspace{-0.1em}S_{\psi }%
\overrightarrow{\partial }{}^{\ast A}S_{\psi }:\delta _{\mu }\phi ^{A}%
\hspace{-0.1em}\equiv \hspace{-0.1em}\phi ^{A}\overleftarrow{s}\hspace{-0.2em%
}\mu ,  \label{softbreakm}
\end{equation}%
where $\overleftarrow{s}$ is the generator of BRST transformations (Slavnov
generator in YM theories). This fact was established in \cite{Reshetnyak}.
In December 2011, one of the authors (A.A.R.) drew the attention of his
coauthors (P.M.~Lavrov) in \cite{llr1} to the research \cite{mandalr} which
attempted to use FD BRST transformations \cite{JM} for relating the vacuum
functionals in YM and Gribov--Zwanziger (GZ) under different gauges. An
explicit calculation of the functional Jacobian for a change of variables
induced by FD BRST transformations in YM theories with a finite parameter $%
\mu $ was made in \cite{LL1}, to establish the gauge-independence of $%
Z_{\psi }$ under a finite change of the gauge, $\psi \rightarrow \psi
+\Delta \psi $, and afterwards in \cite{LL2}, to solve equation (\ref%
{softbreakm}) with $M(\phi )=H(A)$, in a way different from anticanonical
transformations, as compared to \cite{llr1}.

The present article reviews the study of finite BRST and BRST-antiBRST
transformations (including the case of field-dependent parameters) and the
way they influence the properties of the quantum action and path integral in
conventional quantization. The article has the following organization.
Section~\ref{BRSTproposal} presents the definitions of finite BRST and
BRST-antiBRST transformations. The algorithm for calculating the functional
determinants related to these transformations is briefly examined in Section~%
\ref{Jacobian}. The implications of this calculation to the quantum
structure of gauge theories are presented in Section~\ref{Implication}.

We use the DeWitt condensed notation and the conventions of \cite%
{Reshetnyak,MR1}, e.g., we use $\epsilon (F)$ for the value of Grassmann
parity of a quantity $F$. Derivatives with respect to (anti)field variables $%
\phi ^{A},\phi _{A}^{\ast }$ and sources $J_{A}$ are denoted by $%
\overleftarrow{\partial }^{A},(\overrightarrow{\partial }_{A}^{\ast })$ and $%
\overrightarrow{\partial }_{(J)}^{A}$. The raising and lowering of $\mathrm{%
Sp}\left( 2\right) $ indices, $\big(\overleftarrow{s}^{a},\overleftarrow{s}%
_{a}\big)=\big(\varepsilon ^{ab}\overleftarrow{s}_{b},\varepsilon _{ab}%
\overleftarrow{s}{}^{b}\big)$, are carried out by a constant antisymmetric
tensor $\varepsilon ^{ab}$, $\varepsilon ^{ac}\varepsilon _{cb}=\delta
_{b}^{a}$, $\varepsilon ^{12}=1$.

\section{Proposals for Finite BRST Transformations}

\label{BRSTproposal} %%%%%%%%%%%%%%%%%%%%%%%%%%%%%%%%%%%%%%%%%%%%%%%%

The problem of softly broken BRST symmetry (SB BRST) in general gauge
theories was solved in \cite{Reshetnyak} on a basis of finite FD
(\textquotedblleft gauged\textquotedblright\ in the terminology of \cite%
{Reshetnyak}) BRST transformations (invariance transformations for the
integrand in (\ref{genfun}) at $J=M=0$) with finite odd-valued parameters $%
\mu (\phi ,\phi ^{\ast })$ depending on external antifields $\phi _{A}^{\ast
}$, $\epsilon (\phi _{A}^{\ast })+1=\epsilon (\phi ^{A})=\epsilon _{A}$, and
fields $\phi ^{A}$ whose contents include the classical fields $A^{i}$, $%
i=1,..,n$, with gauge transformations $\delta A^{i}=R_{\alpha }^{i}(A)\xi
^{\alpha }$, $\alpha =1,..,m<n$, the ghost, antighost, and
Nakanishi--Lautrup fields $C^{\alpha },\bar{C}^{\alpha },B^{\alpha }$, $%
\epsilon (A^{i},\xi ^{\alpha },C^{\alpha },\bar{C}^{\alpha },B^{\alpha })$= $%
\big({\epsilon }_{i}$, ${\epsilon }_{\alpha }$, ${\epsilon }_{\alpha }+1$, ${%
\epsilon }_{\alpha }+1$, ${\epsilon }_{\alpha }\big)$, as well as the
additional towers of fields depending on the (ir)reducibility of the theory.
The generating functional of Green's functions depending on external sources
$J_{A}$, $\epsilon (J_{A})=\epsilon _{A}$, with an SB BRST symmetry term $M$%
, $\epsilon (M)=0$, is given by%
\begin{equation}
\hspace{-0.7em}Z_{\psi ,M}(J,\phi ^{\ast })\hspace{-0.15em}=\hspace{-0.3em}%
\int \hspace{-0.25em}d\phi \ \exp \left\{ \hspace{-0.25em}\frac{\imath }{%
\hbar }S_{\psi }(\phi ,\phi ^{\ast })\hspace{-0.1em}+\hspace{-0.1em}M(\phi
,\phi ^{\ast })+J_{A}\phi ^{A}\hspace{-0.25em}\right\} \hspace{-0.2em},\,%
\mathrm{with}\,\overleftarrow{s}_{e}\hspace{-0.1em}=\hspace{-0.1em}{%
\overleftarrow{\partial }}_{A}{\overrightarrow{\partial }}{}^{\ast A}S_{\psi
}\equiv {\overleftarrow{\partial }}_{A}S_{\psi }^{\ast A}\hspace{-0.2em},
\label{genfun}
\end{equation}%
where the generator $\overleftarrow{s}_{e}$, $\phi ^{A}\overleftarrow{s}%
_{e}=S_{\psi }^{\ast A}(\phi ,\phi ^{\ast })$, reduces at $\phi ^{\ast }=0$
to the usual generator $\overleftarrow{s}$ of (FD) BRST transformations, $%
\delta _{\mu }\phi ^{A}=S_{\psi }^{\ast A}(\phi ,0)\mu $, and fails to be
nilpotent, due to the quantum master equation for $S_{\psi }$,
\begin{equation}
\left[ \Delta \exp \left\{ \frac{\imath }{\hbar }S_{\psi }\right\}
=0\Longleftrightarrow S_{\psi }{\overleftarrow{\partial }}_{A}{%
\overrightarrow{\partial }}{}^{\ast A}S_{\psi }=\imath \hbar \Delta S_{\psi }%
\right] \Rightarrow (\overleftarrow{s}_{e})^{2}={\overleftarrow{\partial }}%
_{A}\big(S_{\psi }^{\ast A}{\overleftarrow{\partial }}_{B}\big)S_{\psi
}^{\ast B}\neq 0,  \label{qme}
\end{equation}%
with $\Delta =(-1)^{\epsilon _{A}}{\overrightarrow{\partial }}_{A}{%
\overrightarrow{\partial }}{}^{\ast A}$. The Jacobian induced by a change of
variables\footnote{%
In the case $\mu \overleftarrow{s}_{e}\neq 0$, the set $\{g(\mu )\}$, for $%
\phi ^{\prime }=\phi g\left( \mu \right) $, cannot be presented as Lie group
elements: $g(\mu )\neq \exp \left( {\overleftarrow{s}_{e}}{\mu }\right) ${.}}
$\phi ^{A}\rightarrow \phi ^{\prime A}=\phi ^{A}(1+\overleftarrow{s}_{e}\mu
) $ was calculated originally in \cite{Reshetnyak}:
\begin{eqnarray}
\hspace{-0.7em} &\hspace{-0.7em}&\hspace{-0.5em}\mathrm{Sdet}\left\Vert \phi
^{\prime }{}^{A}\overleftarrow{\partial }_{B}\right\Vert =\exp \hspace{-0.1em%
}\left\{ \hspace{-0.1em}\mathrm{Str}\mathrm{ln}\hspace{-0.1em}\left( \hspace{%
-0.1em}\delta _{B}^{A}+(S_{\psi }^{\ast A}\mu )\overleftarrow{\partial }_{B}%
\hspace{-0.1em}\right) \hspace{-0.1em}\right\} =\exp \hspace{-0.1em}\bigg\{%
\hspace{-0.1em}\mathrm{Str}\hspace{-0.15em}\sum_{n=1}\hspace{-0.2em}\frac{%
(-1)^{n+1}}{n}\hspace{-0.2em}\bigg((S_{\psi }^{\ast A}\mu )\overleftarrow{%
\partial }_{B}\hspace{-0.2em}\bigg)^{n}\hspace{-0.1em}\bigg\}  \label{sdet}
\\
\hspace{-0.7em} &\hspace{-0.7em}&\hspace{-0.5em}\Longrightarrow \hspace{%
-0.15em}\sum_{n=1}^{\infty }\hspace{-0.1em}\frac{(-1)^{n+1}}{n}\hspace{-0.1em%
}\mathrm{Str}\big(\hspace{-0.1em}(S_{\Psi }^{\ast A}\mu )\overleftarrow{%
\partial }_{B}\hspace{-0.1em}\big)^{n}\hspace{-0.1em}=\hspace{-0.2em}%
\sum_{n=1}^{\infty }\hspace{-0.15em}\frac{(-1)^{n}}{n}\hspace{-0.1em}(\mu
\overleftarrow{s}_{e})^{n}\hspace{-0.1em}-\hspace{-0.2em}\sum_{n=2}^{\infty }%
\hspace{-0.15em}{(-1)^{n}}\hspace{-0.1em}(\mu \overleftarrow{s}%
_{e})^{n-2}\mu _{A}\bigl(S_{\psi }^{\ast A}\overleftarrow{s}_{e}\bigr)\mu
\label{trJac} \\
\hspace{-0.7em} &\hspace{-0.7em}&\hspace{-0.5em}\ \ +\bigl(\Delta S_{\psi }%
\bigr)\mu \ \ \Longrightarrow \mathrm{Sdet}\left\Vert \phi ^{\prime }{}^{A}%
\overleftarrow{\partial }_{B}\right\Vert \ =\ \big(1+\mu \overleftarrow{s}%
_{e}\big)^{-1}\Big\{1+\overleftarrow{s}_{e}\mu \Bigr\}\Big\{1+\bigl(\Delta
S_{\psi }\bigr)\mu \Bigr\},  \label{jacobianres}
\end{eqnarray}%
under a suitable condition of convergence for the series in (\ref{trJac}),
and reduces, in a rank-$1$ theory with a closed gauge algebra, to the form%
\begin{equation}
\lbrack \Delta S_{\psi },\overleftarrow{s}{}^{2}]=[0,0],\ \mathrm{where}\
\overleftarrow{s}_{e}=\overleftarrow{s}\Longrightarrow \mathrm{Sdet}%
\left\Vert \Phi ^{\prime }{}^{A}\overleftarrow{\partial }_{B}\right\Vert \
=\ \big(1+\mu \overleftarrow{s}\big)^{-1}\ ,  \label{sdetca}
\end{equation}%
which is the same as in YM theories.

The construction of finite BRST-antiBRST Lagrangian transformations solving
the same problem within a suitable quantization scheme (starting from YM
theories), is problematic in view of the BRST-antiBRST non-invariance of the
gauge-fixed quantum action $S_{F}$ in a form more than linear in $\mu _{a}$,
$S_{F}(g_{l}(\mu _{a})\phi )=S_{F}(\phi )+O(\mu _{1}\mu _{2})$, with the
gauge condition encoded by a gauge boson $F(\phi )$. This problem was solved
in January 2014, by constructing finite BRST-antiBRST transformations in a
group form, $\{g(\mu _{a})\}$, using an appropriate set of variables $\Gamma
^{p}$, according to \cite{MR1}%
\begin{equation}
\hspace{-0.1em}\left\{ G\left( g(\mu _{a})\Gamma \right) =G\left( \Gamma
\right) \ \mathrm{and}\ G\overleftarrow{s}{}^{a}=0\right\} \hspace{-0.1em}%
\Rightarrow g\left( \hspace{-0.1em}\mu _{a}\hspace{-0.1em}\right) =1+%
\overleftarrow{s}{}^{a}\mu _{a}+\textstyle\frac{1}{4}\overleftarrow{s}{}^{a}%
\overleftarrow{s}{}_{a}\mu ^{2}=\exp \left\{ \hspace{-0.1em}\overleftarrow{s%
}{}^{a}\mu _{a}\hspace{-0.1em}\right\} ,  \label{bab}
\end{equation}%
for an arbitrary regular functional $G(\Gamma)$, where $\mu ^{2}\equiv \mu _{a}\mu ^{a}$, and $\overleftarrow{s}{}^{a}$, $%
\overleftarrow{s}{}^{2}\equiv \overleftarrow{s}{}^{a}\overleftarrow{s}{}_{a}$
are the generators of BRST-antiBRST and mixed BRST-antiBRST transformations
in the space of $\Gamma ^{p}$. These transformations, however, cannot be
presented as group elements (in terms of an $\exp $-like relation)\ for an $%
\mathrm{Sp}(2)$ doublet $\mu _{a}$ which is not closed under FD
BRST-antiBRST transformations: $\mu _{a}\overleftarrow{s}_{b}\neq 0$.

In YM theories, the quantum action $S_{F}\left( \phi \right) $ in $R_{\xi }$%
-like gauges [given by a bosonic gauge functional $F(\phi )$] and the finite
BRST-antiBRST transformations are constructed using an explicit form of
BRST-antiBRST generators in the space of fields $\phi ^{A}=(A^{i},C^{\alpha
},\bar{C}^{\alpha },B^{\alpha })$, being identical with those of the
Faddeev--Popov quantization rules \cite{FP} and organized in $\mathrm{Sp}(2)$%
-symmetric tensors, $(A^{i},C^{\alpha {}a},B^{\alpha })$ = $\big(A^{\mu
m},C^{ma},B^{m}\big)$, as follows \cite{MR1}:
\begin{eqnarray}
\hspace{-0.5em} &\hspace{-0.5em}&S_{F}(\phi )=S_{0}(A)-\textstyle\frac{1}{2}%
F_{\xi }\overleftarrow{s}{}^{2},\ S_{0}(A)=-\frac{1}{4}\int \hspace{-0.2em}%
d^{D}xG_{\mu \nu }^{m}G^{m\mu \nu },\ G_{\mu \nu }^{m}=\partial _{\lbrack
\mu }A_{\nu ]}^{m}+f^{mnl}A_{\mu }^{n}A_{\nu }^{l},  \label{variat} \\
\hspace{-0.5em} &\hspace{-0.5em}&\ F_{\xi }(\phi )=\textstyle\frac{1}{2}\int
d^{D}x\ \big(- A_{\mu }^{m}A^{m\mu }+\frac{\xi }{2}\varepsilon
_{ab}C^{ma}C^{mb}\big)\ ,  \label{F(A,C)} \\
\hspace{-0.5em} &\hspace{-0.5em}&\Delta A_{\mu }^{m}=D_{\mu }^{mn}C^{na}\mu
_{a}-\textstyle\frac{1}{2}\left( D_{\mu }^{mn}B^{n}+\frac{1}{2}%
f^{mnl}C^{la}D_{\mu }^{nk}C^{kb}\varepsilon _{ba}\right) \mu ^{2}\ ,
\label{DAmm} \\
\hspace{-0.5em} &\hspace{-0.5em}&\Delta B^{m}=-\textstyle\frac{1}{2}\left(
f^{mnl}B^{l}C^{na}+\frac{1}{6}f^{mnl}f^{lrs}C^{sb}C^{ra}C^{nc}\varepsilon
_{cb}\right) \mu _{a}\ ,  \label{DBm} \\
\hspace{-0.5em} &\hspace{-0.5em}&\Delta C^{ma}=\left( \varepsilon ^{ab}B^{m}-%
\textstyle\frac{1}{2}f^{mnl}C^{la}C^{nb}\right) \mu _{b}-\textstyle\frac{1}{2%
}\left( f^{mnl}B^{l}C^{na}+\textstyle\frac{1}{6}%
f^{mnl}f^{lrs}C^{sb}C^{ra}C^{nc}\varepsilon _{cb}\right) \mu ^{2}.
\label{DCma}
\end{eqnarray}%
These relations are inherited from the non-Abelian gauge transformations in
terms of a covariant derivative,
\begin{equation}
\delta A_{\mu }^{m}(x)=D_{\mu }^{mn}(x)\zeta ^{n}(x)=\int d^{D}y\ R_{\mu
}^{mn}(x;y)\zeta ^{n}(y)\ ,\ \ \mathrm{where}\ \ i=(\mu ,m,x),\alpha =(n,y),
\label{R(A)}
\end{equation}%
where the generator of gauge transformations $R_{\mu }^{mn}(x;y)$ leaves the
classical action $S_{0}$ invariant with accuracy up to the first order in
arbitrary functions $\zeta ^{n}$: $\mathcal{S}_{0}(A+\delta A)=\mathcal{S}%
_{0}(A)+o(\zeta ^{n})$; the metric tensor is $\eta _{\mu \nu }=\mathrm{diag}%
(-,+,\ldots ,+)$, and $f^{lmn}$\ are the totally antisymmetric~$su(\hat{N})$
structure constants, $l,m,n=1,\ldots ,\hat{N}{}^{2}-1$. The $N=2$
BRST-antiBRST invariant action for YM theories coincides with the $N=1$ BRST
invariant action only in Landau gauge, $\xi =0$, whereas in the gauges $%
F_{\xi \neq 0}$ including the Feynmann gauge, $\xi =1$, the action $S_{F}$
contains terms quartic in the ghosts $C^{m{}a}$, leading thereby to
additional vertices in the Feynmann integrals. This is a property of
quantization with a global SUSY symmetry higher than $N=1$ BRST symmetry.

In general gauge theories, such as reducible ones (for higher-spin field
theories, see \cite{BuchbinderReshetnyak,Y[2]reshetnyak} and references
therein), or those with an open gauge algebra (as in some SUSY models), the
corresponding space of triplectic variables $\Gamma _{{tr}}^{p}=(\phi
^{A},\phi _{Aa}^{\ast },\bar{\phi}_{A},\pi ^{Aa},\lambda ^{A})$ in the $%
\mathrm{Sp}(2)$-covariant Lagrangian quantization method \cite{BLT1}
contains, in addition to $\phi ^{A}$, three sets of antifields $\phi
_{Aa}^{\ast }$, $\bar{\phi}_{A}$, $\epsilon (\phi _{Aa}^{\ast },\bar{\phi}%
_{A})=(\epsilon _{A}+1,\epsilon _{A})$, as sources to BRST, antiBRST and
mixed BRST-antiBRST transformations, and three sets of Lagrangian
multipliers $\pi ^{Aa},\lambda ^{A}$, $\epsilon (\pi ^{Aa},\lambda
^{A})=(\epsilon _{A}+1,\epsilon _{A})$, introducing the gauge. The
corresponding generating functional of Green's functions, $Z_{F}(J)$,
\begin{equation}
\hspace{-0.5em}Z_{F}(J)=\hspace{-0.2em}\int \hspace{-0.1em}d\Gamma \hspace{%
-0.1em}\exp \hspace{-0.1em}\left\{ \hspace{-0.15em}\big(\imath /\hbar \big)%
\Big[W+\phi _{a}^{\ast }\pi ^{a}+\bar{\phi}\lambda -\textstyle\frac{1}{2}F%
\overleftarrow{U}{}^{2}+J\phi \Big]\hspace{-0.15em}\right\} ,\
\overleftarrow{U}_{a}={\overleftarrow{\partial }}_{A}\pi ^{Aa}+\varepsilon
^{ab}{\overleftarrow{\partial }_{Ab}^{(\pi )}}\lambda ^{A}  \label{z(0)}
\end{equation}%
is invariant, at $J=0$, with respect to finite BRST-antiBRST transformations
(for constant $\mu _{a}$) in the space of $\Gamma _{{tr}}^{p}$, which are
given by (\ref{bab}), however, with the functional $G_{{tr}}=G(\Gamma _{{tr}%
}^{p})$:
\begin{align}
& \hspace{-0.7em}\Gamma _{{tr}}^{p}\rightarrow \Gamma _{{tr}}^{\prime
p}=\Gamma _{{tr}}^{p}\left( 1+\overleftarrow{s}{}^{a}\mu _{a}+\textstyle%
\frac{1}{4}\overleftarrow{s}{}^{2}\mu ^{2}\right) \equiv \Gamma _{{tr}%
}^{p}g(\mu _{a})\Longrightarrow \mathcal{I}_{\Gamma _{{tr}}g(\mu
_{a})}^{\left( F\right) }=\mathcal{I}_{\Gamma _{{tr}}}^{\left( F\right) }\
\mathrm{for}\,Z_{F}=\int \mathcal{I}_{\Gamma _{{tr}}}^{\left( F\right) },
\label{Gamma_fin} \\
& \hspace{-0.7em}\mathrm{where}\ \overleftarrow{s}{}^{a}\hspace{-0.15em}=%
\hspace{-0.1em}\Big(\hspace{-0.15em}{\overleftarrow{\partial }}_{A},{%
\overleftarrow{\partial }}{}_{(\phi ^{\ast })}^{Aa},{\overleftarrow{\partial
}}_{(\bar{\phi})}^{A},{\overleftarrow{\partial }}_{Ab}^{(\pi )}\hspace{-0.1em%
}\Big)\hspace{-0.1em}\Big(\hspace{-0.1em}\pi ^{Aa},W_{,A}(-1)^{\epsilon
_{A}},\varepsilon ^{ab}\phi _{Ab}^{\ast }(-1)^{\epsilon _{A}+1},\varepsilon
^{ab}\lambda ^{A}\hspace{-0.1em}\Big)^{T}\hspace{-0.4em},\{\overleftarrow{s}{%
}^{a},\overleftarrow{s}{}^{b}\}\hspace{-0.1em}\neq \hspace{-0.1em}0,  \notag
\\
& \hspace{-0.7em}\ \mathrm{provided\ that}\ \left( \Delta ^{a}+(\imath
/\hbar )\varepsilon ^{ab}\phi _{Ab}^{\ast }{\overrightarrow{\partial }}_{(%
\bar{\phi})}^{A}\right) \exp \left\{ \frac{\imath }{\hbar }W\right\} =0,\
\mathrm{for}\ \Delta ^{a}=(-1)^{\epsilon _{A}}{\overrightarrow{\partial }}%
_{A}{\overrightarrow{\partial }}{}^{\ast Aa},  \label{eqme}
\end{align}%
with respective classical action $S_0(A)$ being be the boundary condition for the equation (\ref{eqme}) on the bosonic functional $W$: $W(\phi, \phi
_{a}^{\ast }, \bar{\phi})\big|_{\phi
_{a}^{\ast }= \bar{\phi}=0}=S_0(A)$.  
The restricted generators $\overleftarrow{U}{}^{a}=\left. \overleftarrow{s}{}%
^{a}\right\vert _{\phi ,\pi ,\lambda }$ are nilpotent and satisfy the
algebra $\{\overleftarrow{U}{}^{a},\overleftarrow{U}{}^{b}\}=0$.

%%%%%%%%%%%%%%%%%%%%%%%%%%%%%%%%%%%%%%%%%%%%%%%%%%%%%%%%%%%%%%%%%

\section{Jacobians of Finite $N=1,2$ BRST Transformations}

\label{Jacobian}

The Jacobian (\ref{jacobianres}) allows one to solve the problem of SB BRST
symmetry in general gauge theories \cite{Reshetnyak} and was examined in
detail \cite{MR4} for an equivalent representation of $Z_{\psi ,M}(J,\phi
^{\ast })$ in (\ref{genfunbv}), as well as for BRST transformations in an
extended space which contains, besides $\phi ^{A}$, also internal (included
in the path integral) antifields $\tilde{\phi}_{A}^{\ast }$ and Lagrangian
multipliers ${\lambda }^{A}$ to Abelian hypergauge conditions, $\mathcal{G}%
_{A}(\phi ,\phi ^{\ast })=\phi _{A}^{\ast }-\psi (\phi )\overleftarrow{%
\partial }_{A}$:
\begin{eqnarray}
\hspace{-0.5em} &\hspace{-0.5em}&Z_{\psi ,M}(J,\phi ^{\ast })=\int d{\Gamma }%
\exp \left\{ \frac{\imath }{\hbar }S({\phi },\tilde{\phi}{}^{\ast })+%
\mathcal{G}_{A}\big({\phi },\tilde{\phi}^{\ast \prime }+\phi ^{\ast }\big)%
\lambda ^{A}+M({\phi },{\phi }{}^{\ast })+J\phi \right\} ,  \label{genfunbv}
\\
\hspace{-0.5em} &\hspace{-0.5em}&\mathrm{with}\ {\Gamma }^{p}\rightarrow {%
\Gamma }^{p\prime }={\Gamma }^{p}\big(1+\overleftarrow{{s}}\mu \big),\
\mathrm{where}\ {\Gamma }^{p}\overleftarrow{{s}}=\big(\phi ^{A},\tilde{\phi}%
_{A}^{\ast },\lambda ^{A}\big)\overleftarrow{{s}}=\big(\lambda ^{A},S%
\overleftarrow{\partial }_{A},0\big) \\
\hspace{-0.5em} &\hspace{-0.5em}&\ \mathrm{and}\ \mathrm{Sdet}\left\Vert
\Gamma ^{p\prime }\overleftarrow{\partial }_{q}^{\Gamma }\right\Vert =\big(%
1+\mu \overleftarrow{{s}}\big)^{-1}\Big\{1+\bigl(\Delta S \bigr)\mu %
\Bigr\}+O(\mu \overleftarrow{{s}}\mu ),  \label{fdbrst}
\end{eqnarray}%
with a field-dependent $\mu (\Gamma )$ in (\ref{fdbrst}), which reduces (for $\phi^*$-independent  $\mu (\Gamma )$: $\mu (\Gamma )\overleftarrow{\partial }^{\ast A}=0$) to the
Jacobian of \cite{brstbv}, i.e., one without $O(\mu \overleftarrow{{s}}\mu )$%
. The bosonic functional $S(\phi ,\phi ^{\ast })$ in (\ref{genfunbv}) is a
proper solution of (\ref{qme}), with the classical action $S_{0}$ as the
boundary condition at $\phi ^{\ast }=0$.

For BRST-antiBRST transformations in YM theories, the technique of
calculating the Jacobian was first examined for functionally-dependent
parameters $\mu _{a}=\Lambda (\phi )\overleftarrow{s}_{a}$ with an
even-valued functional $\Lambda $ and was developed in \cite{MR1}, resulting
in, $\phi ^{\prime }{}^{A}\equiv \phi ^{{}}{}^{A}g(\Lambda (\phi )%
\overleftarrow{s}_{a})$,
\begin{eqnarray}
&&\ J_{\Lambda (\phi )\overleftarrow{s}_{a}}=\mathrm{Sdet}\left\Vert \phi
^{\prime }{}^{A}\overleftarrow{\partial }_{B}\right\Vert =\exp \hspace{-0.1em%
}\left\{ \hspace{-0.1em}\mathrm{Str}\,\mathrm{ln}\left( \delta
_{B}^{A}+M_{B}^{A}\right) \right\} ,\ \ \mathrm{for}\ \
M_{B}^{A}=P_{B}^{A}+Q_{B}^{A}+R_{B}^{A}  \label{jacobianN2YM} \\
&&\ =\phi ^{A}\overleftarrow{s}^{a}(\mu _{a}\overleftarrow{\partial }%
_{B})+\mu _{a}\big[(\phi ^{A}\overleftarrow{s}^{a})\overleftarrow{\partial }%
_{B}-\textstyle\frac{1}{2}(\phi ^{A}\overleftarrow{s}^{2})(\mu ^{a}%
\overleftarrow{\partial }_{B})\big](-1)^{\epsilon _{A}+1}+\textstyle\frac{1}{%
4}\mu ^{2}(\phi ^{A}\overleftarrow{s}^{2}\overleftarrow{\partial }_{B}),
\notag \\
&&\ \mathrm{Str}(P+Q+R)^{n}=\mathrm{Str}(P+Q)^{n}+C_{n}^{1}\mathrm{Str}%
P^{n-1}R,\ \ \mathrm{for}\ \ C_{n}^{k}=n!/k!(n-k)!,
\label{jacobianN2YMrules} \\
&&\ \mathrm{Str}(P+Q)^{n}=\left\{
\begin{array}{l}
\mathrm{Str}P^{n}+n\mathrm{Str}P^{n-1}Q+C_{n}^{2}\mathrm{Str}P^{n-2}Q^{2},\
n=2,3, \\
\mathrm{Str}P^{n}+n\textstyle\sum_{k=0}^{2}\mathrm{Str}P^{n-k}Q^{k}+K_{n}%
\mathrm{Str}P^{n-3}QPQ,\ n>3%
\end{array}%
\right.   \label{P+Q} \\
&&\ \Longrightarrow J_{\Lambda (\phi )\overleftarrow{s}_{a}}=\exp \hspace{%
-0.1em}\left\{ \hspace{-0.1em}\,\sum_{n=1}(-1)^{n-1}n^{-1}\mathrm{Str}%
(P_{B}^{A})^{n}\right\} =\left( 1-\textstyle\frac{1}{2}\Lambda
\overleftarrow{s}{}^{2}\right) ^{-2},  \label{jacobianN2YMres}
\end{eqnarray}%
where $K_{n}=\left[ \frac{n+1}{2}-2\right] C_{n}^{1}+\big((n+1)\,\mathrm{mod}%
\,2\big)C_{\left[ \frac{n}{2}\right] }^{1}$, with $[x]$ being the integer
part of $x\in \mathbb{R}$. The Jacobian (\ref{jacobianN2YMres}) cannot be
derived from the Jacobian (\ref{sdetca}) corresponding to FD BRST
transformations in YM theories.\footnote{%
The corresponding Jacobian is not equal to the expression $J=$ $\mathrm{Sdet}%
\left\Vert \phi ^{A\prime }\overleftarrow{\partial }_{B}\right\Vert =\big(1+%
\big[\bar{\mu}\overleftarrow{\bar{s}}\mu \big]\overleftarrow{s}\big)^{-1}=1-%
\bar{\mu}\overleftarrow{\bar{s}}\mu \overleftarrow{s}+o(\mu \bar{\mu}),$
which is presented in [http://theor.jinr.ru/sqs15/Lavrov.pdf]. The just
mentioned expression does not allow one to control the gauge independence of
$Z_{F}$ in (\ref{z(0)}) for infinitesimal FD parameters $(\mu _{1},\mu _{2})=
$ $(\mu ,\bar{\mu})$, $(\overleftarrow{s}^{1},\overleftarrow{s}^{2})$ = $(%
\hat{s},\hat{\bar{s}})$, either in YM or in general gauge theories.} For
functionally-independent FD parameters $\mu _{a}(\phi )\neq \Lambda
\overleftarrow{s}{}_{a}$, the above algorithm (\ref{jacobianN2YM})--(\ref%
{jacobianN2YMres}) of calculating $J_{\mu _{a}}$ involves a generalization
of (\ref{P+Q}) to the case $P^{n}\not\equiv f^{n-1}P$, examined separately
for odd and even $n>3$, which leads to \cite{MR5}%
\begin{equation}
\hspace{-0.5em}\hspace{-0.5em}J_{\mu _{a}}=\exp \hspace{-0.1em}\left\{
\hspace{-0.1em}\,\mathrm{tr}\sum_{n=1}(-1)^{n-1}n^{-1}\mathrm{Str}%
(P_{B}^{A})^{n}\right\} =\exp \hspace{-0.1em}\left\{ \hspace{-0.1em}\,-%
\mathrm{tr}\ln (e+m)\right\} ,\ m_{b}^{a}=\mu _{b}\overleftarrow{s}{}^{a},
\label{jacobianN2YMgen}
\end{equation}%
where $\left( e\right) _{b}^{a}$ and $\mathrm{tr}$ denote $\delta _{b}^{a}$
and trace over $\mathrm{Sp}(2)$ indices. The Jacobian (\ref{jacobianN2YMgen}%
) is generally not BRST-antiBRST exact; however, it reduces at $\mu
_{a}=\Lambda \overleftarrow{s}_{a}$ to the Jacobian (\ref{jacobianN2YMres}),
due to
\begin{equation}
\mathrm{tr}\,m_{b}^{a}=\mathrm{tr}\,\Lambda \overleftarrow{s}{}_{b}%
\overleftarrow{s}{}^{a}=-(1/2)\mathrm{tr}\,\delta _{b}^{a}\Lambda
\overleftarrow{s}{}^{2}\Rightarrow \mathrm{tr}\,m^{n}=2[-(1/2)\Lambda
\overleftarrow{s}{}^{2}]^{n}\Rightarrow J_{\mu _{a}}=J_{\Lambda
\overleftarrow{s}_{a}}.  \label{corrfd}
\end{equation}

In general gauge theories (\ref{z(0)})--(\ref{eqme}), the calculation of
Jacobians induced by FD BRST-antiBRST transformations was first carried out
in \cite{MR2,MR4} with functionally-dependent parameters $\mu _{a}=\Lambda
(\phi ,\pi ,\lambda )\overleftarrow{U}_{a}$ and then in \cite{MR5} with
arbitrary parameters $\mu _{a}(\Gamma _{tr})$, including
functionally-independent $\mu _{a}(\phi ,\pi ,\lambda )$. The result is
given by
\begin{align}
& \hspace{-0.7em}J_{\Lambda \overleftarrow{U}_{a}}\hspace{-0.1em}=\hspace{%
-0.1em}\mathrm{Sdet}\left\Vert \left[ {\Gamma _{tr}^{p}}{g}(\Lambda
\overleftarrow{U}_{a})\right] \overleftarrow{\partial }{}_{q}^{\Gamma
}\right\Vert \hspace{-0.1em}=\hspace{-0.1em}\exp \left[ -\left( \Delta
^{a}W\right) \mu _{a}-\textstyle\frac{1}{4}\left( \Delta ^{a}W\right)
\overleftarrow{s}_{a}\mu ^{2}\right] \left( 1-\textstyle\frac{1}{2}\Lambda
\overleftarrow{s}{}^{2}\right) ^{-2}\hspace{-0.1em},  \label{superJaux6} \\
& \hspace{-0.7em}J_{\mu _{a}(\phi ,\pi ,\lambda )}=\exp \Big\{-\left( \Delta
^{a}W\right) \mu _{a}-\textstyle\frac{1}{4}\left( \Delta ^{a}W\right) {%
\overleftarrow{s}}_{a}\mu ^{2}-\mathrm{tr}\,\ln \left( e+{m}\right) \Big\},
\label{J-arb} \\
& \hspace{-0.7em}\ J_{\mu _{a}(\Gamma _{tr})}=J|_{\mu _{a}(\phi ,\pi
,\lambda )\rightarrow \mu _{a}(\Gamma _{tr})}\exp \Big\{\textstyle\frac{1}{4}%
(\mu _{a}\overleftarrow{\partial }{}_{p}^{\Gamma })\left[ (e+{m})^{-1}\right]
_{b}^{a}\left( {\Gamma }_{tr}^{p}\overleftarrow{s}{}^{2}\overleftarrow{s}{}%
^{b}\right) \mu ^{2}\Big\}.  \label{J-arb-gen}
\end{align}%
The second multiplier in (\ref{J-arb-gen}) draws a difference between the
Jacobians $J_{\mu _{a}(\phi ,\pi ,\lambda )}$ and $J_{\mu _{a}(\Gamma
_{tr})} $, because $\overleftarrow{s}_{a}$ are not reduced to the nilpotent $%
\overleftarrow{U}_{a}$ as they act on $\Gamma _{tr}^{p}$. This result
generalizes the Jacobian for $\mu _{a}(\phi ,\pi ,\lambda )$ obtained by a
special Green function, using a $t$-parametric rescaling of the Lie
equations in \cite{brstbv}, and cannot be achieved by the prescription
therein, due to the non-integrability condition $\overleftarrow{s}_{a}%
\overleftarrow{s}_{b}\overleftarrow{s}_{c}\neq 0$. For constant parameters $%
\mu _{a}$, the Jacobians (\ref{superJaux6})--(\ref{J-arb-gen}) are reduced
to the same expression, being an $\hbar $-deformation of the classical
master equations (\ref{eqme}) and their differential consequences obtained
by applying $\overleftarrow{s}_{a}$.

In generalized Hamiltonian formalism, the Jacobians of corresponding FD
BRST-antiBRST transformations were calculated from first principles by the
rules (\ref{jacobianN2YM})--(\ref{jacobianN2YMgen}) in \cite{MR3,MR5},
whereas the so-called \emph{linearized }finite BRST-antiBRST transformations
$\Gamma \rightarrow \Gamma ^{\prime }=\Gamma (1+\overleftarrow{s}^{a}\mu
_{a})$ in YM and arbitrary first-class constraint dynamical systems, along
with a calculation of Jacobians and a discussion of their implications on
the quantum theory, were presented in \cite{MR5}.

\section{Implications of Finite BRST Transformations}

\label{Implication} The proposals for $N=1$ and $N=2$ finite BRST
transformations allow one to establish (in the case of constant $\mu $ and $%
\mu _{a}$) the finite BRST (identical with the case of small $\mu $) and
BRST-antiBRST invariance of the integrand in (\ref{genfunbv}) with a
vanishing $N=1,2$ SB BRST symmetry term $M$, as well as in (\ref{z(0)}) for
YM and general gauge theories.

For FD parameters, finite BRST transformations allow one to obtain a new
form of the Ward identity and to establish the gauge independence of the
path integral under a finite change of the gauge, $\psi \rightarrow \psi
+\psi ^{\prime }$, provided that the SB BRST symmetry term $M=M_{\psi }$
transforms to $M_{\psi +\psi ^{\prime }}=M_{\psi }(1+\overleftarrow{s}\mu
(\psi ^{\prime }))$, with $\mu (\psi ^{\prime })$ being a solution of a
so-called compensation equation, as one makes a change of variables
corresponding to an FD BRST transformation in the integrand of $Z_{\psi
,M_{\psi }}(0,\phi ^{\ast })$:
\begin{equation}
Z_{\psi ,M_{\psi }}(0,\phi ^{\ast })=Z_{\psi +\psi ^{\prime },M_{\psi +\psi
^{\prime }}}(0,\phi ^{\ast })\Rightarrow \psi ^{\prime }(\phi ,\lambda |\mu
)=\textstyle\frac{\hbar }{i}\left[ \sum\nolimits_{n=1}\textstyle\frac{%
(-1)^{n-1}}{n}\left( \mu \overleftarrow{s}\right) {}^{n-1}\right] \mu .
\label{comeq}
\end{equation}%
The Ward identity, depending on the FD parameter $\mu (\psi ^{\prime })=-%
\frac{i}{\hbar }g(y)\psi ^{\prime }$ , for $g(y)=1-\exp \{y\}/y$, $y\equiv ({%
i}/{\hbar })\psi ^{\prime }\overleftarrow{s}$, and the gauge-dependence
problem are described by the respective expressions \cite{MR4}
\begin{equation}
\hspace{-0.7em}\left\langle \hspace{-0.3em}\left\{ \hspace{-0.2em}1+\frac{i}{%
\hbar }\left[ \hspace{-0.1em}J_{A}\phi ^{A}+M_{\psi }\hspace{-0.1em}\right]
\overleftarrow{s}\mu (\psi ^{\prime })\hspace{-0.2em}\right\} \left( \hspace{%
-0.1em}1+\mu (\psi ^{\prime })\overleftarrow{s}\hspace{-0.1em}\right) {}^{-1}%
\hspace{-0.3em}\right\rangle _{\psi ,M,J}\hspace{-0.4em}=1\,\mathrm{and}%
\,\left\langle \hspace{-0.15em}(J_{A}\phi ^{A}+M_{\psi })\overleftarrow{s}%
\hspace{-0.25em}\right\rangle _{\psi ,M,J}\hspace{-0.3em}=0,  \label{mWIbvbr}
\end{equation}%
as one makes averaging with respect to $Z_{\psi ,M_{\psi }}(J,\phi ^{\ast })$%
. The above equations are equivalent to those of \cite{Reshetnyak,llr1}

In turn, FD BRST-antiBRST transformations solve the same problem (allowing
for the presence of an SB BRST-antiBRST symmetry term $M=M_{F}$ and not
touching upon the unitarity issue \cite{MR4,MR6}) under a finite change of
the gauge, $F\rightarrow F+F^{\prime }$, provided that the term $M_{F}$
transforms to $M_{F+F^{\prime }}=M_{F}(1+\overleftarrow{s}^{a}\mu
_{a}(F^{\prime })+\frac{1}{4}\overleftarrow{s}^{2}\mu {}^{2}(F^{\prime }))$,
with $\mu _{a}(F^{\prime };\phi ,\pi ,\lambda )=\Lambda \overleftarrow{U}%
_{a} $ being a solution to the corresponding compensation equation, as one
makes a change of variables corresponding to an FD BRST transformation in
the integrand of $Z_{F}(0)$ from (\ref{z(0)}):
\begin{equation}
Z_{F}(0)=Z_{F+F^{\prime }}(0)\Rightarrow F^{\prime }(\phi ,\pi ,\lambda |\mu
_{a})=4\imath {\hbar }\left[ \sum\nolimits_{n=1}\textstyle\frac{(-1)^{n-1}}{%
2^{n}n}\left( \Lambda \overleftarrow{U}^{2}\right) {}^{n-1}\Lambda \right] .
\label{comeqsp}
\end{equation}%
As a result, the corresponding Ward identity, with the FD parameters $\mu
_{a}(F^{\prime })=\frac{i}{2\hbar }g(y)F^{\prime }\overleftarrow{U}_{a}$, $%
\Lambda (\Gamma |{F}^{\prime })=\frac{i}{2\hbar }g(y){F}^{\prime }$, for $%
y\equiv ({i}/{4\hbar })F^{\prime }\overleftarrow{U}{}^{2}$, and the
gauge-dependence problem acquire the form \cite{MR4}
\begin{eqnarray}
&&\textstyle\left\langle \left\{ 1+\frac{i}{\hbar }J_{A}\phi ^{A}\left[
\overleftarrow{U}^{a}\mu _{a}(\Lambda )+\frac{1}{4}\overleftarrow{U}^{2}\mu
^{2}(\Lambda )\right] -\frac{1}{4}\left( \frac{i}{\hbar }\right)
{}^{2}J_{A}\phi ^{A}\overleftarrow{U}^{a}J_{B}(\phi ^{B})\overleftarrow{U}%
_{a}\mu ^{2}(\Lambda )\right\} \right.  \label{mWI} \\
&&\quad \left. \times \textstyle\left( 1-\frac{1}{2}\Lambda \overleftarrow{U}%
^{2}\right) {}^{-2}\right\rangle _{F,J}=1,  \notag \\
&&Z_{F+F^{\prime }}(J)=Z_{F}(J)\left\{ 1+\textstyle\left\langle \frac{i}{%
\hbar }J_{A}\phi ^{A}\left[ \overleftarrow{U}^{a}\mu _{a}\left( \Gamma |-{F}%
^{\prime }\right) +\frac{1}{4}\overleftarrow{U}^{2}\mu ^{2}\left( \Gamma |-{F%
}^{\prime }\right) \right] \right. \right.  \notag \\
&&\ -\left. \left. \textstyle(-1)^{\varepsilon _{B}}\left( \frac{i}{2\hbar }%
\right) ^{2}J_{B}J_{A}\left( \phi ^{A}\overleftarrow{U}{}^{a}\right) \left(
\phi ^{B}\overleftarrow{U}_{a}\right) \mu ^{2}\left( \Gamma |-{F}^{\prime
}\right) \right\rangle _{F,J}\right\} ,  \label{GDInew}
\end{eqnarray}%
with source-dependent average expectation value with respect to $Z_{F}(J)$
corresponding to a gauge-fixing $F(\phi )$.

By choosing the $N=1$ or $N=2$ SB BRST symmetry term $M(\phi )$ as the
horizon functional $H(A)$ in Landau (or Coulomb) gauge, and assuming the
gauge independence of the path integrals $Z_{H,\psi }$, $Z_{H,F}$ under a
finite change of the gauge condition, $\psi \rightarrow \psi +\psi ^{\prime
} $ or $F\rightarrow F+F^{\prime }$, one can determine the functional $H(A)$
in a new reference frame, $\psi +\psi ^{\prime }$ or $F+F^{\prime }$, of the
respective $N=1,2$ BRST symmetry setting, with account taken of (\ref{comeq}%
), (\ref{comeqsp}):%
\begin{equation}
H_{\psi ^{\prime }}(\phi )=H(A)\left\{ 1+\overleftarrow{s}\mu (\psi ^{\prime
})\right\} \ \mathrm{or}\ H_{F^{\prime }}(\phi )=H(A)\left\{ 1+%
\overleftarrow{s}{}^{a}\mu _{a}(F^{\prime })+\frac{1}{4}\overleftarrow{s}{}%
^{2}\mu ^{2}(F^{\prime })\right\} .  \label{DeltaH}
\end{equation}

Notice in conclusion that the above $N=1,2$ FD BRST transformations make it
possible to study their influence on the Yang--Mills, Gribov--Zwanziger,
Freedman--Townsend models, and the Standard Model, as well as on the concept
of average effective action \cite{Reshetnyak,MR1,MR2,MR4,MR5}. The case of
functionally-dependent parameters $\mu _{a}\left( \Gamma \right) =\Lambda
\left( \Gamma \right) \overleftarrow{s}^{a}+\psi _{a}\left( \Gamma \right) $
with a vanishing BRST-antiBRST \textquotedblleft
divergence\textquotedblright , $\psi _{a}\overleftarrow{s}^{a}=0$, was
examined in Sec.~4.1. of Ref. \cite{MR5} and implies a modified form of the
compensation equation, due to a nontrivial contribution of $\psi _{a}$ to
the corresponding Jacobian.

\paragraph{Acknowledgments}

The authors are grateful to the organizers of the International Workshop
SQS'15 for kind hospitality, as well as to M.O. Katanaev and K.V.
Stepanyantz for their interest and discussions. The study was supported by
the RFBR grant No. 16-42-700702, and was also partially supported by the
Tomsk State University Competitiveness Improvement Program.

\end{document}